\documentclass[12pt]{article}
\usepackage{epsfig}
\usepackage{amssymb}

\def\l{\left(}
\def\r{\right)}

\newcommand{\be}{\begin{equation}}
\newcommand{\ee}{\end{equation}}

\newcommand{\bg}{\begin{gather}}
\newcommand{\eg}{\end{gather}}

\def\gtap{\raisebox{-.55ex}{\rlap{$\sim$}} \raisebox{.4ex}{$>$}}
\def\gsim{\mathrel{\gtap}}

\begin{document}

\title{Violation of Lorentz Invariance and neutral component of UHECR}
\author{S.L.~Dubovsky$^{a}$ and P.G.~Tinyakov$^{b,a}$\\
 $^a${\small\it Institute for Nuclear Research, Moscow 117312, Russia
 }\\ $^b${\small\it Institute of Theoretical Physics, University of
 Lausanne,} \\ {\small\it CH-1015 Lausanne, Switzerland} }
\date{}
\maketitle

\begin{abstract}
The observed clustering of ultra-high energy cosmic rays suggests the
existence of a neutral component. The models with violation of Lorentz
invariance may explain this component by neutrons becoming stable
above some threshold energy $E_0$. The protons, in turn, may become
unstable above some energy $E_1>E_0$. We calculate the dependence of
the threshold energies $E_0$ and $E_1$ on the parameters of the model
and find $E_1/E_0\gsim 1.5$. We argue that the characteristic
threshold behavior of charged and neutral components may be used as
the specific signature of models with violation of Lorentz
invariance. The existence of the neutron stability threshold $E_0$ can
be investigated with already existing data.
\end{abstract}

\section{Introduction} 
Ultra-high energy cosmic ray (UHECR) experiments are believed to be
one of the promising places where physics beyond the Standard Model
may be discovered. Until recently this expectation was based on a
number of arguments of which the non-observation \cite{observations}
of the GZK cutoff \cite{GZK} is most acknowledged. The case is
strengthened by the absence of nearby candidate sources in the
directions of the observed UHECR, large-scale isotropy of arrival
directions and generic difficulties with acceleration of particles to
energies of order $10^{20}$~eV.

The apparent absence of the GZK cutoff by itself can be explained
within the conventional physics. For instance, several models have
been proposed which assume isotropization of arrival directions in
strong magnetic fields and attribute the observed UHECR to neutron
stars of our Galaxy \cite{olinto}, Virgo cluster \cite{virgo} or the
nearby radio-galaxy Cen~A \cite{strongMF}. Regardless of a particular
model, assuming hard injection spectrum in combination with local
overdensity of sources at scales $\sim 50$~Mpc makes the discrepancy
between the observed and predicted spectra significantly smaller
\cite{olinto1,medina}.

A new ingredient in this problem is the small scale anisotropy of
UHECR. The study of arrival directions of UHECR reveals the existence
of clusters of events \cite{clusters1,Takeda:1999sg,Uchihori:2000gu}.
The analysis based on angular correlation function shows that the
typical size of clusters is comparable with the experimental angular
resolution \cite{Tinyakov:2001ic,takeda2}. A most natural
interpretation of this result is the existence of point-like sources
of UHECR whose image is smeared by errors in determination of UHECR
arrival directions.

If this interpretation is believed, it has several important
consequences. First, it excludes the models which assume diffuse
propagation of UHECR in strong magnetic fields. Second, it makes very
unlikely the scenarios which explain the absence of the GZK cutoff by
assuming hard spectrum and local overdensity of sources. Indeed, in
the latter case the arguments based on statistics of clustering impose
unrealistic bounds on the local density of sources
\cite{Dubovsky:2000gv,Fodor:2001yi}. Finally, the observed clustering
suggests the existence of neutral primary particles, since otherwise
they would be deflected in the Galactic magnetic fields and clusters
would not be so tight.

Even stronger argument in favor of neutral particles follows from the
correlation of arrival directions with BL Lacertae objects (BL Lac)
\cite{Tinyakov:2001nr}. BL Lacs are located at cosmological
distances. For instance, two of the BL Lacs which coincide with
triplets of UHECR are both at $\sim 600$~Mpc. If primary particles of
the triplet events indeed came from BL Lacs, they must be neutral;
moreover, they must be able to propagate over cosmological distances
without substantial attenuation.

The existence of neutral particles imposes further constraints on
possible models of UHECR. Within the Standard Model, there are two
stable neutral particles, photon and neutrino. The photon attenuation
length due to $e^+e^-$ pair production on infrared background is of
order $10-20$~Mpc in the energy range $10^{19}-10^{20}$~eV
\cite{sigl}, so photons cannot explain the observed events unless the
infrared background is extremely low and/or their initial energies are
extremely high.

Neutrinos by themselves cannot be particles which initiate airshowers
if standard neutrino cross sections are assumed. They have to be
converted into hadrons and photons through interactions with
primordial neutrino background
\cite{Weiler:1999sh,Fargion:1999ft,Gelmini:1999qa}. Photons obtained
in this way could account for the observed tight clusters. There are
generic problems with this scenario \cite{Yoshida:1998it}. First, the
smallness of neutrino cross sections implies large neutrino
flux. Second, the production of neutrinos by accelerated protons
requires significantly higher proton energy, while even energies of
order $10^{20}$~eV are difficult to achieve in most acceleration
sites. If these difficulties can be overcome, neutrino may be an
appealing candidate.

While the situation with neutrinos is unclear, it is worth considering
other possibilities. Outside of the Standard Model, several candidates
for neutral primary particles have been proposed. One of them is a
light SUSY hadron \cite{Chung:1998rz,Albuquerque:1999va} which can be,
e.g., a $uds$-gluino bound state or gluon-gluino bound state. Another
possible candidate suggested recently is a light sgoldstino
\cite{Gorbunov:2001gc}. The models of this type have an advantage 
of explaining the neutral component and the absence of the GZK cutoff
at the same time. The disadvantage is that primary particles have to
be produced by accelerated protons, which leads to losses in both flux
and energy. As a result, already very tight requirements on
acceleration energy and efficiency become even tighter.

An attractive model which shares the same advantages but does not have
the latter drawback has been proposed in
Ref.~\cite{Coleman:1999ti}. It is based on possible violation of
Lorentz invariance at high energies \cite{Coleman:1999ti,Colladay}
(see, e.g., \cite{bertolami} for further discussion of Lorentz
invariance violation in the context of UHECR). Models of this type can
be constructed without giving up the main principles of quantum field
theory in the framework of the brane world scenario
\cite{Visser,Csaki:2000dm,Dubovsky:2001fj}.
Under rather general assumptions, violation of Lorentz invariance can
be described phenomenologically by introducing different maximum
velocities for different particles. As noted in
Ref.~\cite{Coleman:1999ti}, these parameters can be arranged in such a
way that the main cause of the GZK cutoff, the pion photoproduction
process, is not operative. At the same value of parameters, neutron
may become heavier than proton at sufficiently high energy, and
therefore stable. Thus, it can serve as a neutral primary particle
which is indistinguishable from proton in other respects. Moreover,
protons (which are stable at low energies) may be accelerated in a
usual way. Once they reach the threshold energy and become unstable,
they decay into neutrons by ``$\beta$-decay'' with no loss in the flux
and practically no loss in energy. Another way to produce UHE neutrons
is in collisions of UHE protons with synchrotron photons in the
acceleration site.

How the above models can be distinguished? A clear signature of the
neutrino model is the direct detection of ultra-high energy neutrinos,
whose flux is expected to be in the range of sensitivity of the Pierre
Auger and Telescope Array experiments \cite{neutrflux}. An indirect
argument in favor of neutrino models would be identification of the
neutral component as photons. On the contrary, the hadronic neutral
component speaks in favor of light SUSY hadron models or violation of
Lorentz invariance.

The purpose of this paper is to point out the signature which can
discriminate between the latter two possibilities. It is based on the
fact that within the model of Ref.~\cite{Coleman:1999ti} the behavior
of proton and neutron masses with energy follows a certain pattern:
with increasing energy, first the neutron becomes stable at some
energy $E_0$, and then proton becomes unstable at some higher energy
$E_1>E_0$. In other words, there is a window in which both neutron and
proton are stable, while at higher energies proton becomes
unstable. Thus, in the model of Ref.~\cite{Coleman:1999ti}, one
generically expects that there is only charged component at $E<E_0$,
both charged and neutral components at $E_0<E<E_1$, and only neutral
component at $E>E_1$. The appearance and disappearance of components
is a threshold effect and is, therefore, a step-like function of
energy. This step-like behavior is the specific signature of models
with violation of Lorentz invariance.

\section{Kinematics of the nucleon decays}
Let us now calculate the width of the energy window in which the two
components coexist. Following Ref.~\cite{Coleman:1999ti} consider the
model characterized by the maximum attainable velocities of neutron,
proton, electron and neutrino equal to $c_n$, $c_p$, $c_e$ and
$c_{\nu}$, respectively. Without loss of generality one may set
\[ 
c_n=1\;.
\]   
Certainly, $c_p$, $c_e$ and $c_{\nu}$ should be very close to 1 as
well. The differences between them are at the level of $\lesssim
10^{-20}$~\cite{Coleman:1999ti}. As it was shown in
Ref.~\cite{Coleman:1999ti}, the neutron is stable at high energies
only if
\[
c_p,\; c_e,\; c_{\nu}\geq 1\;.
\]
Moreover, if $c_p\neq 1$ then proton becomes unstable at high enough
energy.  We will show that the ratio $E_0/E_1$ is determined by the
parameter
\begin{equation}
\alpha \equiv {c^2_e-c^2_n\over c^2_p-c^2_n}, 
\label{alpha}
\end{equation}
while the absolute magnitudes of both $E_0$ and $E_1$ are proportional
to $\epsilon^{-1}$, where 
\begin{equation}
\epsilon \equiv (c_p^2-c_n^2)^{1/2}. 
\label{epsilon}
\end{equation}

Following the formalism of Ref.~\cite{Coleman:1999ti}, let us find the
energy $E_0$ above which the neutron is stable, as a function of the
parameters $c_p$, $c_e$ and $c_{\nu}$. In order to do this one has to
find the minimum possible energy $E_{\rm mn}(p_n)$ of the potential
decay products, i.e. proton, electron and neutrino, at fixed total
momentum $p_n$. The condition of stability of the neutron with the
momentum $p_n$ and energy $E_n=\sqrt{p_n^2+m_n^2}$ has the form
\begin{equation}
\label{neutronstab}
\sqrt{ p_n^2+m_n^2}\leq E_{\rm mn}(p_n)\;.
\end{equation}
As it was shown in Ref.~\cite{Coleman:1999ti}, the equality in
Eq.~(\ref{neutronstab}) may hold only at one value of the momentum
$p_n$. Consequently, the energy $E_0$ is equal to
\begin{equation}
E_0=\sqrt{p_0^2+m_n^2}\;,
\label{E0}
\end{equation} 
where $p_0$ is a solution to the equation
\be
\label{0}
\sqrt{p_0^2+m_n^2}=E_{mn}(p_0)\;.
\ee
Clearly, the minimum total energy of the proton, electron and neutrino
with the fixed total momentum is achieved when the momenta of all
three particles are collinear. Thus, in order to calculate
$E_{mn}(p_0)$ one should find the minimum value of the following
function
\be
\label{fneut}
{\cal E}_n(p_p,p_e,p_{\nu})=
c_p\sqrt{p_p^2+m_p^2c_p^2}+c_e\sqrt{p_e^2+m_e^2c_e^2}+c_{\nu}p_{\nu} 
\ee 
of three positive variables $p_p,p_e$ and $p_{\nu}$ subject to the
constraint
\be
\label{constr}
p_0=p_p+p_e+p_{\nu}\;.
\ee
This minimum may be either an extremum of the function ${\cal
E}_n(p_p,p_e,p_{\nu})$, or belong to the boundary of the triangular
region determined by the constraint (\ref{constr}) and positivity
conditions
\[
p_p,\;p_e,\;p_{\nu}\geq 0\;.
\]
As is shown in the Appendix, the minimum is achieved at zero neutrino
energy and momentum,
\[
p_{\nu}=0\;.
\]
Therefore, in order to find the energy at which neutron becomes stable
one should study the kinematics of two-body decay $n\to pe$. As a
consequence, $E_0$ does not depend on the neutrino velocity
$c_{\nu}$. The problem reduces to the following system of equations
\begin{eqnarray}
\label{ntope1}
p_0&=&p_e+p_p\\
\label{ntope2}
{c_e p_e\over \sqrt{p_e^2+m_e^2c_e^2}}&=&{c_p p_p\over
\sqrt{p_p^2+m_p^2c_p^2}}\\ 
\label{ntope3}
\sqrt{p_0^2+m_n^2}&=&c_e\sqrt{p_e^2+m_e^2c_e^2}+c_p\sqrt{p_p^2+m_p^2c_p^2}\;.
\end{eqnarray}
Here eq.(\ref{ntope1}) is the momentum conservation condition
(\ref{constr}) with $p_{\nu}=0$, eq.(\ref{ntope2}) is the extremality
condition for the function ${\cal E}(p_p,p_e,0)$ and eq.(\ref{ntope2})
is the stability condition (\ref0). It is straightforward to solve
this system analytically in the case $c_e=c_p$ with the result
\be
\label{col1}
E_0\approx\sqrt{m_n^2-(m_p+m_e)^2\over c_p^2-1}\sim 3.85\times
10^{19} \l {10^{-12}\over \epsilon}\r\;\mbox{eV},
\ee
in agreement with Ref.~\cite{Coleman:1999ti}.  

\begin{figure}[nt]
\begin{center}
\begin{picture}(380,300)(0,0)
\put(0,0){
\epsfig{file=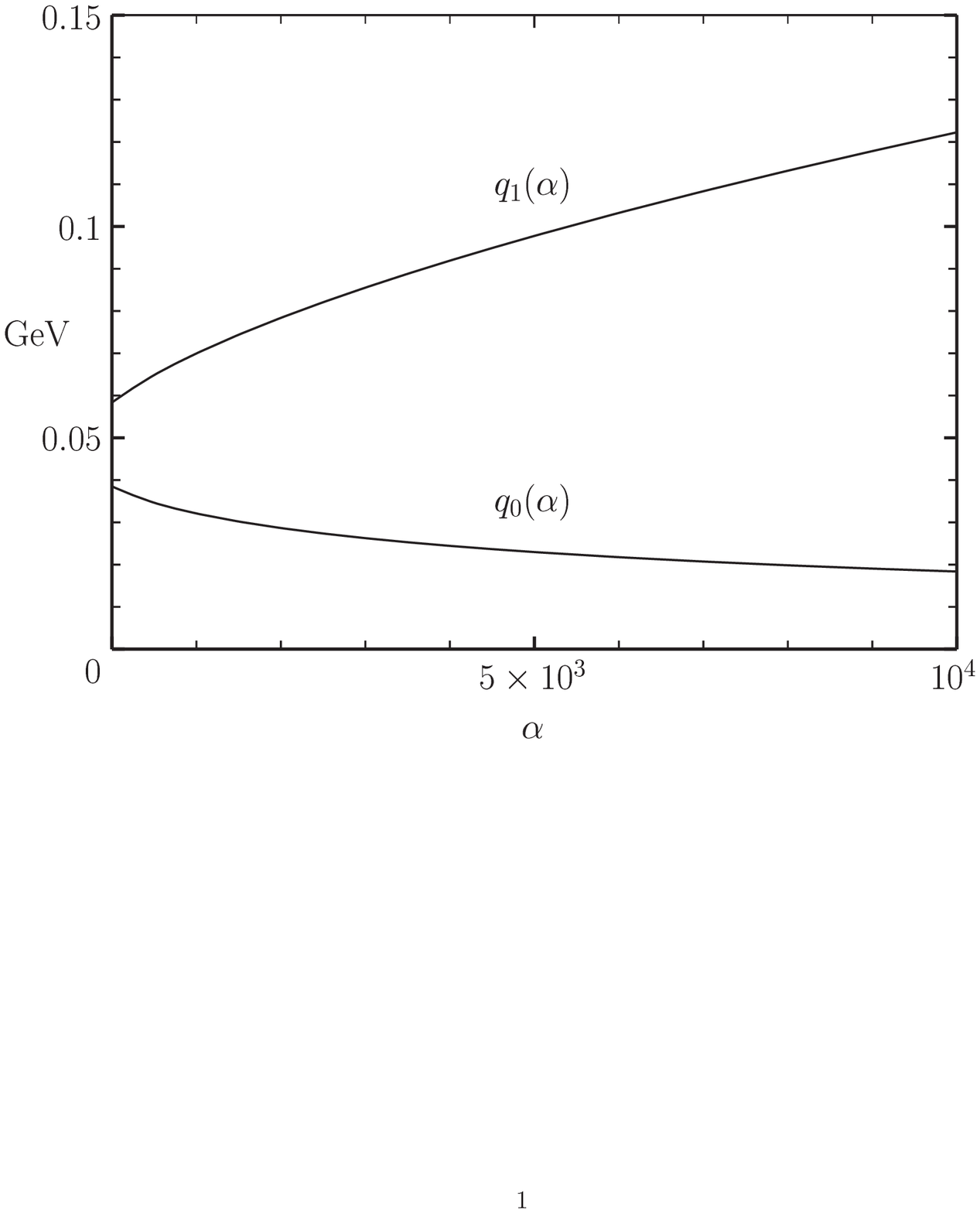,%
bbllx=35pt,bblly=360pt,%
bburx=525pt,bbury=715pt,%
width=330pt,height=240pt,%
clip=}}
\end{picture}
\end{center}
\caption{The dependence of the rescaled threshold momenta
$q_0\approx\epsilon E_0$ and $q_1\approx\epsilon E_1$ on the
parameter $\alpha$.}
\label{fig:1}
\end{figure}

In the case $c_p\neq c_e$ it is convenient to express the velocities
in terms of the parameters $\alpha$ and $\epsilon$ according to
\[
c_p^2=1+\epsilon^2,\;c_e^2=1+\alpha\epsilon^2,
\]
and rescale the momenta 
\[
p_i={q_i\over \epsilon}.
\]
We assume $\epsilon$ to be very small, but make no assumptions about
the value of $\alpha$. Expanding eqs.(\ref{ntope1})-(\ref{ntope3}) to
the lowest non-trivial order in $\epsilon$ we obtain a system of
polynomial equations for three unknowns $q_0,\;q_p,\;q_e$, which
depends on one parameter $\alpha$. This system can be solved for $q_0$
numerically. The threshold energy is determined by the relation 
\begin{equation}
E_0 = {q_0(\alpha)\over \epsilon},
\label{Et0}
\end{equation}
where the difference between $E_0$ and $p_0$ has been neglected. The
numerical solution for $q_0$ as a function of $\alpha$ is plotted in
Fig.~\ref{fig:1}.

Let us now find the threshold energy $E_1$ above which the decay $p\to
ne\bar\nu$ is kinematicaly allowed. We follow the similar procedure as
in the case of neutron, {\it i.e.}, find the threshold momentum $p_1$
from the equation
\be
\label{eqforpr}
c_p\sqrt{p_1^2+m_p^2c_p^2}=E_{mp}(p_1)\;,
\ee
where $E_{mp}(p_p)$ is the minimum total energy of a neutron, electron and
neutrino with the fixed total momentum $p_p$. This energy is 
obtained by the minimization of the function 
\be
\label{fprot}
{\cal E}_p(p_n,p_e,p_{\nu})=\sqrt{p_n^2+m_n^2}
+c_e\sqrt{p_e^2+m_e^2c_e^2}+c_{\nu}p_{\nu} 
\ee
under the constraints
\be
\label{pconstr}
p_n+p_e+p_{\nu}=p_p,\qquad p_n,p_e,p_{\nu}\geq 0\;.
\ee
As in the case of a neutron one may show that the minimum is reached
at $p_{\nu}=0$ and therefore it is sufficient to study the kinematics
of the two-body decay $p\to ne$ when both momenta $p_p$ and $p_e$ are
non-zero. The system of equations determining the threshold momentum
$p_1$ has the following form (cf. Eqs.~(\ref{ntope1})-(\ref{ntope3}))
\begin{eqnarray}
\label{tone1}
p_1&=&p_e+p_n\\
\label{tone2}
{c_e p_e\over \sqrt{p_e^2+m_e^2c_e^2}}&=&{p_n\over
\sqrt{p_n^2+m_n^2}}\\ 
\label{tone3}
c_p\sqrt{p_1^2+c_p^2m_p^2}&=&c_e\sqrt{p_e^2+c_e^2m_e^2}+
\sqrt{p_n^2+m_n^2}\;.
\end{eqnarray}
Again, the solution is straightforward to find in the case $c_p=c_{e}$
(cf. Ref.~\cite{Coleman:1999ti}),
\be
\label{col2}
E_1\approx{\sqrt{m_n^2-(m_p-m_e)^2}\over \epsilon}\sim 5.83\times
10^{19} \l {10^{-12}\over \epsilon}\r \mbox{eV} ,
\ee
At $c_p\neq c_e$, the solution can be obtained numerically. The
resulting function $q_1(\alpha)$ is plotted in Fig.~\ref{fig:1}. At
small $\epsilon$, the threshold energy $E_1$ is related to
$q_1(\alpha)$ by
\begin{equation}
E_1 = {q_1(\alpha)\over \epsilon}. 
\label{E1}
\end{equation}

\begin{figure}[nt]
\begin{center}
\begin{picture}(380,300)(0,0)
\put(0,0){
\epsfig{file=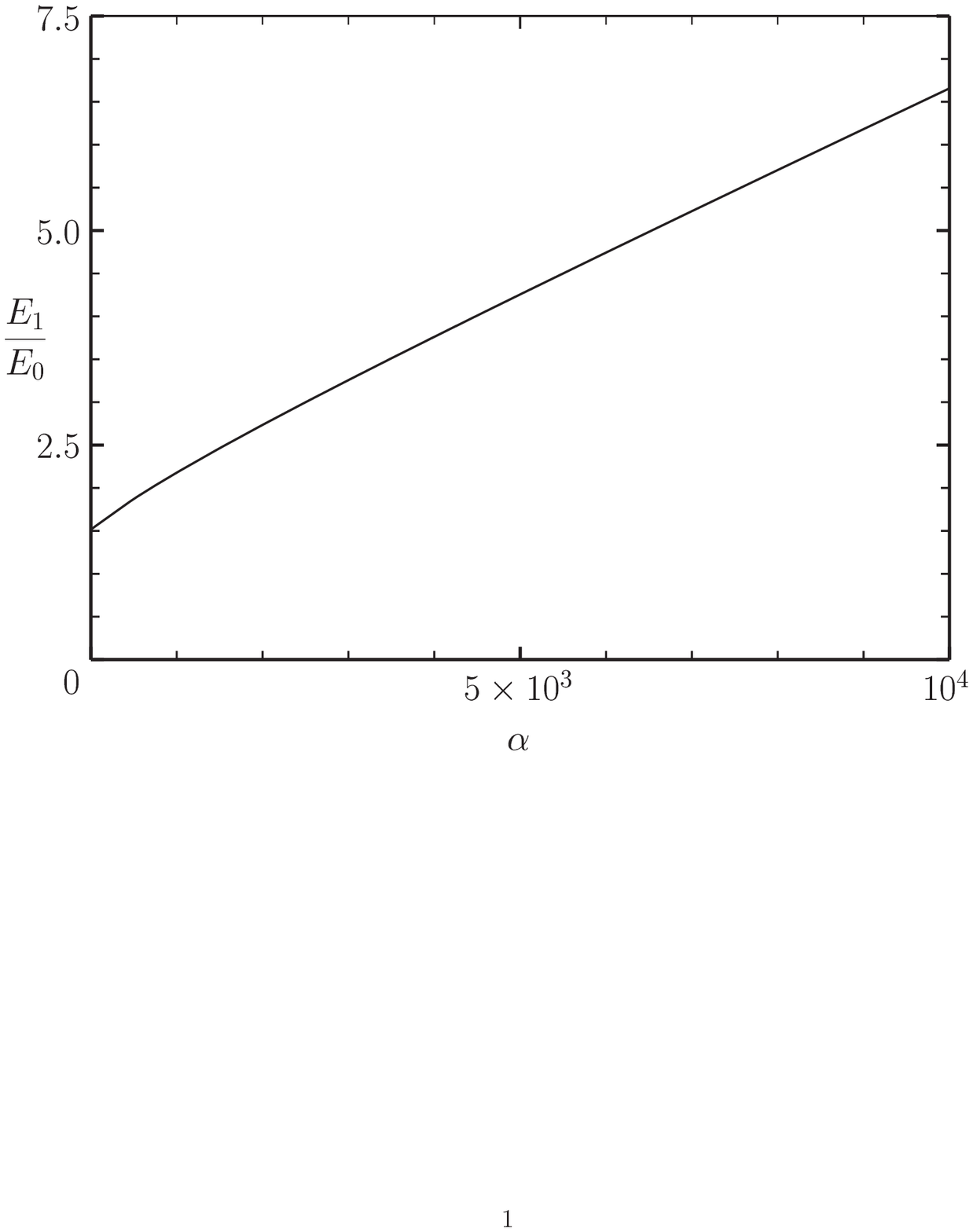,%
bbllx=65pt,bblly=360pt,%
bburx=525pt,bbury=715pt,%
width=320pt,height=240pt,%
clip=}}
\end{picture}
\end{center}
\caption{The ratio of threshold energies $E_1/E_0$ as the function of
the parameter $\alpha$.}
\label{fig:2}
\end{figure}

Comparing Eqs.(\ref{col1}) and (\ref{col2}) one finds that the ratio
$E_1/E_0 \sim 1.5$ at $c_p=c_{e}$ (i.e., at $\alpha=1$). It grows with
$\alpha$, as is shown in Fig.~\ref{fig:2}. The dependence becomes
nearly linear at the values $\alpha\gsim m_p/m_e\sim 2000$.

\section{Discussion} 

As we have seen in Sect.~2, the threshold energies of neutron
stability and proton instability, $E_0$ and $E_1$, depend on two
parameters $\epsilon$ and $\alpha$ as defined by eqs.~(\ref{Et0}) and
(\ref{E1}). The parameter $\epsilon$ sets an overall scale of
$E_0$ and $E_1$, while the parameter $\alpha$ uniquely determines the
ratio $E_1/E_0$. This ratio is always larger than $\sim 1.5$. 

In order to be of phenomenological interest, the neutron stability
threshold $E_0$ must be low enough. If clustering in Yakutsk data 
\cite{Tinyakov:2001ic} is attributed to neutrons, the existence of
triplet with energies $(2.5;2.8;3.4)\times 10^{19}$~eV implies
$E_0<2.5\times 10^{19}$~eV. At $\alpha \ll m_p/m_e$ this requires in
turn $\epsilon > 1.5\times 10^{-12}$, or $c_p^2-c_n^2>2\times
10^{-24}$ (this constraint is somewhat weaker for large values of
$\alpha$, see fig.\ref{fig:1}). This constraint does not contradict
the existing upper bounds \cite{Coleman:1999ti} on $c_p^2-1$ and
$c_n^2-1$. Interestingly, the threshold at $E_0$ may, in principle, be
detected (or ruled out) with already existing experimental data at low
energies $E<4\times 10^{19}$~eV. For this analysis the key issue is
the angular resolution of the experiment, so the AGASA data seems the
most suitable choice. Note that the improvement of upper limits on
$c_p^2-c_n^2$ may rule out the models which explain the neutral
component of UHECR by the violation of the Lorentz invariance.

The threshold of proton instability $E_1$ is more difficult to detect
since this would require good identification of charged component.  In
principle, this can be done by making use of the deflections in the
Galactic and extragalactic magnetic fields. Two different situations
should be distinguished. If at energies smaller than $E_1$ the
extragalactic fields cause random deflections by angles larger than
the typical distance between the sources, the charged component would
look like a uniform background. It will be difficult, if possible at
all, to separate such a component from unresolved sources.

On the contrary, if random deflections are smaller than typical
separation between the sources, one would see ``halos'' of the events
formed by charged primary particles around clusters of the events with
neutral primaries. If correlations with BL Lacs at large angles
\cite{Tinyakov:2001nr} are believed, current data favor the latter
situation. If so, with enough statistics one will be able to identify
the charged component reliably. Even better situation is possible if
random deflections are negligible. In that case not only the charged
particles can be identified, but their charges and the structure of
magnetic fields can, in principle, be determined.

It is worth noting that models with exotic neutral primaries also
generically predict the flux of charged particles, the direct protons
from the source. The difference with the previous case is that in
these models the proton component has a conventional GZK cutoff which
occurs at fixed and calculable energy $E_{\rm GZK}$, while the proton
instability threshold $E_1$ is, in general, different. If by chance
$E_1\simeq E_{\rm GZK}$, the discrimination between the two models
would be very difficult.

To summarize, we argue that the characteristic threshold behavior of
charged and neutral components in models with violation of Lorentz
invariance can be used as their experimental signature. Upper bound on
the neutron stability threshold $E_0$ can be obtained from already
existing data. the proton stability threshold $E_1$ may be seen by the
Pierre Auger experiment \cite{auger} where about 1000 events in the
energy range $10^{19}-10^{20}$~eV are expected in a year.

\section*{Acknowledgements} 

The authors are grateful to M.~Teshima and V.~Rubakov for useful
discussions. This work is supported by the Swiss Science Foundation,
grant 21-58947.99. The work of S.D is supported in part by RFBR grant
99-01-18410, by the Council for Presidential Grants and State Support
of Leading Scientific Schools, grant 00-15-96626, by Swiss Science
Foundation grant 7SUPJ062239 and by CRDF grant (award RP1-2103).
S.D. thanks IPT, University of Lausanne, for hospitality.

\section*{Appendix. Minimization of ${\cal E}_n(p_p,p_e,p_{\nu})$}
\label{app}
It is straightforward to check that function ${\cal
E}_n(p_p,p_e,p_{\nu})$ has an extremum inside the triangular region if
the inequality $c_p,\;c_e>c_{\nu}$ and the following condition is
satisfied
\be
\label{insidecond}
p_n>{c_{\nu}c_e\over(c_e^2-c_{\nu}^2)^{1/2}}m_e
+{c_{\nu}c_p\over(c_p^2-c_{\nu}^2)^{1/2}}m_p\;.
\ee
The total energy of the proton, electron and neutrino in this extremum
is
\be
E_1(p_n)=c_{\nu}p_n+m_pc_p(c_p^2-c_{\nu}^2)^{1/2}+
m_ec_e(c_e^2-c_{\nu}^2)^{1/2}\;. 
\ee
Now one can readily check that the  equation
\[
\sqrt{p_n^2+m_n^2}=E_1(p_n)
\]
is incompatible with the inequality (\ref{insidecond}). Therefore, in
order to determine the threshold momentum $p_0$ of neutron stability
one should find the minimum value of the function ${\cal
E}_n(p_p,p_e,p_{\nu})$ on the boundary of momentum region. For that
one should consider six cases when either one or two of the momenta
$p_p,\;p_e,\;p_{\nu}$ are zero.

Let us start with the case $p_e=0$. The function ${\cal
E}_n(p_p,0,p_{\nu})$ has an extremum with respect to $p_p$ and
$p_{\nu}$ provided that $c_p>c_{\nu}$ and
\[
p_n>{c_{\nu}c_p\over(c_p^2-c_{\nu}^2)^{1/2}}m_p\;.
\]
The energy corresponding to this extremum is
\[
E_2(p_n)=c_{\nu}p_n+m_pc_p(c_p^2-c_{\nu}^2)^{1/2}+
m_ec_e^2\;.
\]
This energy becomes larger than the energy of neutron with the
momentum $p_n$ at very low momenta $p_n\gtrsim
(m_n^2-m_e^2)/2m_e$. Thus, this kinematical channel of the neutron
decay is closed at energies higher than a few GeV even in the
Lorentz-invariant case. It is straightforward to check that the same
happens in the case $p_p=0$ and when any two momenta of the decay
products are zero. Therefore, the minimum of interest corresponds to
the case $p_{\nu}=0$.

The threshold of proton instability is found in a similar way. The
result is the following: in order to find the energy $E_1$, where
proton becomes unstable one should set $p_{\nu}=0$ and study two-body
decay $p\to n,e$. To illustrate some differences with the case of
neutron decay, let us consider just one alternative kinematical
regime, corresponding to point on the boundary of the region
determined by (\ref{pconstr}) where only the proton momentum is
non-zero. In this case equation (\ref{eqforpr}) reduces to
\[
c_p\sqrt{p_1^2+m_p^2c_p^2}=\sqrt{p_n^2+m_n^2}+m_e^2c_e^4
\]
which gives 
\[
p_1\approx{2m_e\over c_p^2-1}\;
\]
This value is indeed much larger than the one in the extremum with
non-zero $p_e$, see Eq.~(\ref{col2}) and Fig.~\ref{fig:1}.


\begin{thebibliography}{99}
\bibitem{observations} M.~Takeda {\it et al.}
Phys. Rev. Lett. {\bf 81} (1998) 1163, astro-ph/9807193;
D.~Bird {\it et al.} Ap.\ J.\ {\bf 441} (1995) 144.

\bibitem{GZK} K.~Greisen, Phys. Rev. Lett. {\bf 16} (1966) 748;
G.T.~Zatsepin and V.A.~Kuzmin, Pisma Zh. Eksp. Teor. Fiz.{\bf 4}
(1966) 144.

\bibitem{olinto} P.~Blasi, R.I.~Epstein, A.V.~Olinto, 
Astrophys.J. 533 (2000) L123, [astro-ph/9912240].

\bibitem{virgo} E-J.~Ahn, G.~Medina-Tanco, P.L.~Biermann, T.~Stanev,
[astro-ph/9911123].

\bibitem{strongMF}  G.R.~Farrar, T.~Piran, Phys. Rev. Lett. {\bf 84}
(2000) 3527, [astro-ph/0010370].

\bibitem{olinto1} M.~Blanton, P.~Blasi, A.V.~Olinto, 
Astropart.Phys. 15 (2001) 275-286, [astro-ph/0009466].

\bibitem{medina} G.A.~Medina-Tanco, Astrophys. J. {\bf 510} (1999) 91.

\bibitem{clusters1} 
X. Chi et al., J. Phys. {\bf G18} (1992) 539;
N. N. Efimov and A. A. Mikhailov, Astropart. Phys. {\bf 2} (1994) 329.

\bibitem{Takeda:1999sg}
M.~Takeda {\it et al.},
Ap. J. ,{\bf 522} (1999) 225,
astro-ph/9902239.

\bibitem{Uchihori:2000gu}
Y.~Uchihori {\it et al.},
Astropart.\ Phys.\ {\bf 13} (2000) 151
[astro-ph/9908193].

\bibitem{Tinyakov:2001ic}
P.~G.~Tinyakov and I.~I.~Tkachev,
astro-ph/0102101.

\bibitem{takeda2} M.~Teshima, Talk at the XI-th 
International School "Particles and Cosmology", April 18 - 24 , 2001,
Baksan Valley, Russia

\bibitem{Dubovsky:2000gv}
S.~L.~Dubovsky, P.~G.~Tinyakov and I.~I.~Tkachev,
Phys.\ Rev.\ Lett.\  {\bf 85} (2000) 1154
[astro-ph/0001317].

\bibitem{Fodor:2001yi}
Z.~Fodor and S.~D.~Katz,
Phys.\ Rev.\ D {\bf 63} (2001) 023002
[hep-ph/0007158].

\bibitem{Tinyakov:2001nr}
P.~G.~Tinyakov and I.~I.~Tkachev,
astro-ph/0102476.

\bibitem{sigl} 
P.~Bhattacharjee, G.~Sigl, Phys.Rept. {\bf 327} (2000) 109. 

\bibitem{Weiler:1999sh}
T.~J.~Weiler,
Astropart.\ Phys.\ {\bf 11} (1999) 303 [hep-ph/9710431].

\bibitem{Fargion:1999ft}
D.~Fargion, B.~Mele and A.~Salis,
Astrophys.\ J.\ {\bf 517} (1999) 725 [astro-ph/9710029].

\bibitem{Gelmini:1999qa}
G.~Gelmini and A.~Kusenko,
Phys.\ Rev.\ Lett.\ {\bf 82} (1999) 5202 [hep-ph/9902354].

\bibitem{Yoshida:1998it}
S.~Yoshida, G.~Sigl and S.~Lee,
Phys.\ Rev.\ Lett.\ {\bf 81} (1998) 5505 [hep-ph/9808324].

\bibitem{Chung:1998rz}
D.~J.~Chung, G.~R.~Farrar and E.~W.~Kolb,
Phys.\ Rev.\ D {\bf 57} (1998) 4606
[astro-ph/9707036].

\bibitem{Albuquerque:1999va}
I.~F.~Albuquerque, G.~R.~Farrar and E.~W.~Kolb,
Phys.\ Rev.\ D {\bf 59} (1999) 015021
[hep-ph/9805288].

\bibitem{Gorbunov:2001gc}
D.~S.~Gorbunov, G.~G.~Raffelt and D.~V.~Semikoz,
hep-ph/0103175.

\bibitem{Coleman:1999ti}
S.~Coleman and S.~L.~Glashow,
Phys.\ Rev.\ D {\bf 59}, 116008 (1999)
[hep-ph/9812418].

\bibitem{Colladay} D.~Colladay, A.~Kostelecky, Phys.Rev. D58 (1998)
116002, [hep-ph/9809521].


\bibitem{bertolami} O. Bertolami, C.S. Carvalho, 
Phys. Rev. {\bf D61} (2000) 103002, [gr-qc/9912117].

\bibitem{Visser} M.~Visser, Phys. Lett. {\bf B159} (1985) 22.

\bibitem{Csaki:2000dm}
C.~Csaki, J.~Erlich and C.~Grojean, Nucl. Phys. {\bf B604} (2001) 312,
[hep-th/0012143].

\bibitem{Dubovsky:2001fj}
S.~L.~Dubovsky,
{\it Tunneling into extra dimension and high-energy 
violation of Lorentz  invariance,}
hep-th/0103205.

\bibitem{neutrflux}  Z. Fodor, S.D. Katz, A. Ringwald, {\it 
Possible detection of relic neutrinos and determination of their mass:
quantitative analysis}, [hep-ph/0105336].

\bibitem{auger} http://auger.cnrs.fr\\
http://www-ta.icrr.u-tokyo.ac.jp

\end{thebibliography}
\end{document}